\documentclass{elsart}

\usepackage{graphicx,natbib,amssymb,lineno,hyperref}
\usepackage{lscape}

\def\astrobj#1{#1}

\newcommand\apj{{ApJ}}%
\newcommand\aj{{AJ}}%

\begin{document}

\begin{frontmatter}
\title{New eccentric eclipsing binary in triple system: SY Phe}

\author{P. Zasche}
\ead{zasche@sirrah.troja.mff.cuni.cz}

\address{Astronomical Institute, Faculty of Mathematics and Physics,
 Charles University Prague, CZ-180 00 Praha 8, V Hole\v{s}ovi\v{c}k\'ach 2, Czech Republic}

\begin{abstract}
Analyzing available photometry from Hipparcos, ASAS, Pi of the sky and Super WASP, we found that
the system SY Phe is a detached eclipsing binary with similar components and orbital period about
5.27089~day. It has a slightly eccentric orbit, however the apsidal motion is probably very slow.
The system undergoes an additional photometric variation on longer time scales superimposed on the
eclipsing light curve. It also contains one distant component, hence the third light was also
considered.
\end{abstract}


\end{frontmatter}

\section{Introduction}

The system SY Phe (= HD 9283 = HIP 7024) was discovered as a variable by \cite{Hoff1949}. Later,
\cite{1958VeSon...3..333H} noted that the Algol-type curve is of BO~Cep-type, and the additional
variability is discussed. The Hipparcos satellite \citep{HIP} reveals two clearly-shaped eclipses
and found the period about 5.27140~day. The photometric amplitude is about 0.5~mag. It is therefore
remarkable that SY~Phe is still classified as a "Variable Star with rapid variations", according to
Simbad, or GCVS \citep{GCVS}.

Spectral type of the system was firstly derived as F8 by \cite{Spencer}, while
\cite{1958VeSon...3..333H} gave the type F4. Later \cite{1978mcts.book.....H} noted the type
F3/F5V. According to the Tycho data \citep{1997A&A...323L..57H} the photometric index is $B_T-V_T =
0.512$~mag, and \cite{2007A&A...474..653V} derived the parallax of the system $\pi = 4.69 \pm
1.49$~mas.

However, the system consists of two visual components separated about 4$^{\prime\prime}$ on the
sky. According to the Washington Double Star Catalog (hereafter
WDS\footnote{\href{http://ad.usno.navy.mil/wds/}{http://ad.usno.navy.mil/wds/}}, \citealt{WDS}),
the astrometric observations of this double do not show any significant change of the position
angle. Therefore, the pair is only weakly gravitationally bounded and its semi-major axis is rather
large. The Hipparcos observations indicate that the eclipsing variable is the A component (the
brighter one).

\begin{figure}
 \includegraphics[width=14cm]{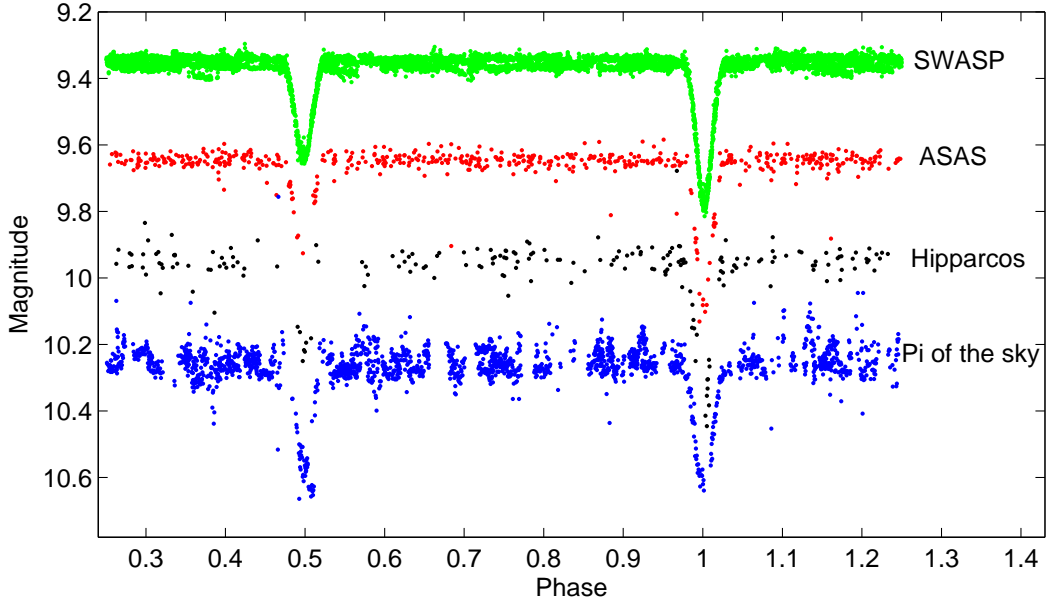}
 \caption{Available light curves of SY Phe.}
 \label{FigLCs}
\end{figure}

\section{Period analysis of the system}

\begin{figure}[b]
 \includegraphics[width=14cm]{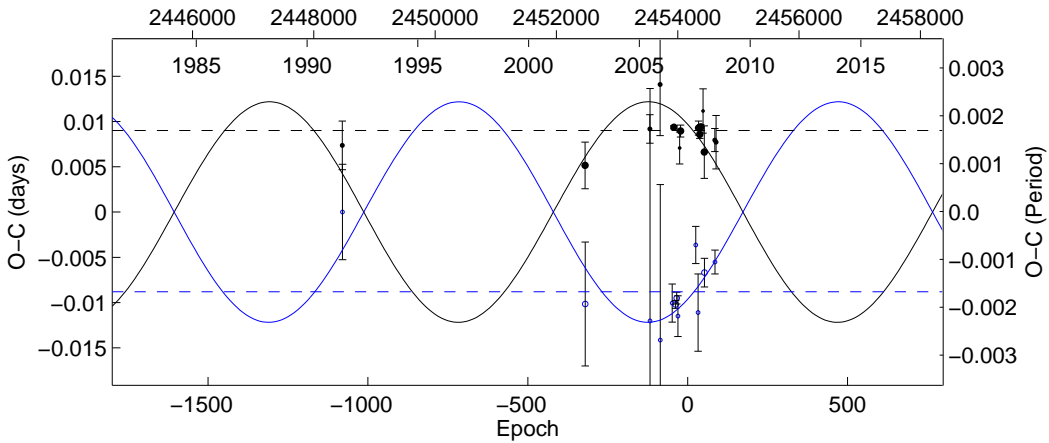}
 \caption{O-C diagram of times of minima derived from available photometry. The black points stand for the
 primary minima, while the blue open circles stand for the secondary ones. The two different apsidal motion
 solutions are plotted as solid curves (Solution I), and dashed curves (II).}
 \label{FigOC}
\end{figure}

The system was observed by several automatic and robotized telescopes and surveys. Besides the old
Hipparcos data, there exist also some photometry of SY Phe obtained by the ASAS survey
\citep{2002AcA....52..397P}, "Pi of the sky" \citep{2005NewA...10..409B}, and Super WASP
\citep{2006PASP..118.1407P}. For the photometry see Fig \ref{FigLCs}. All of these observations
were used for deriving the times of minima of SY~Phe. Both primary and secondary minima were
derived, see Table \ref{TableMin}. Some of the minima have relatively large uncertainty, however if
we plot these minima times in the $O-C$ diagram (see Fig. \ref{FigOC}), there is clearly seen the
eccentric orbit of the system. Primary and secondary minima are well-separated, but the analysis of
apsidal motion is still difficult.

\begin{table}
 \caption{The heliocentric minima times used for the analysis.}
 \label{TableMin} \centering \scalebox{0.80}{
\begin{tabular}{ c c c c c}
\hline \hline
  HJD        & Error  & Type & Filter & Source \\
 2400000+    & [days] &      &        &        \\ \hline
 48470.71478 & 0.00269& Prim &  Hp    & Hipparcos  \\
 48473.34287 & 0.00527& Sec  &  Hp    & Hipparcos  \\
 52471.31528 & 0.00258& Prim &  V     & ASAS       \\
 52473.93542 & 0.00685& Sec  &  V     & ASAS       \\
 53541.30923 & 0.00158& Prim &  V     & ASAS           \\
 53543.92347 & 0.02567& Sec  &  V     & ASAS           \\
 53709.98250 & 0.00564& Prim &  -     & Pi of the sky  \\
 53712.58972 & 0.01718& Sec  &  -     & Pi of the sky  \\
 53907.61660 & 0.00210& Sec  &  SWASP & SWASP          \\
 53936.62590 & 0.00017& Prim &  SWASP & SWASP          \\
 53965.59620 & 0.00044& Sec  &  SWASP & SWASP          \\
 53981.40956 & 0.00065& Sec  &  SWASP & SWASP          \\
 54002.49111 & 0.00225& Sec  &  SWASP & SWASP          \\
 54031.49955 & 0.00176& Prim &  SWASP & SWASP          \\
 54047.31409 & 0.00066& Prim &  SWASP & SWASP          \\
 54297.66860 & 0.00205& Sec  &  SWASP & SWASP          \\
 54334.55734 & 0.00427& Sec  &  SWASP & SWASP          \\
 54342.48405 & 0.00078& Prim &  SWASP & SWASP          \\
 54363.56689 & 0.00044& Prim &  SWASP & SWASP          \\
 54379.38040 & 0.00023& Prim &  SWASP & SWASP          \\
 54416.27834 & 0.00245& Prim &  SWASP & SWASP          \\
 54437.35735 & 0.00290& Prim &  SWASP & SWASP          \\
 54439.97948 & 0.00158& Sec  &  SWASP & SWASP          \\
 54611.29793 & 0.00127& Prim &  -     & Pi of the sky  \\
 54613.91990 & 0.00132& Sec  &  -     & Pi of the sky  \\
 54632.38124 & 0.00295& Prim &  V     & ASAS           \\
 \hline
\end{tabular}}
\end{table}

The method of apsidal motion analysis was described elsewhere, e.g. Gim{\'e}nez \&
Garc{\'{\i}}a-Pelayo (1983), \cite{1995Ap&SS.226...99G}. All of the minima times given in Table
\ref{TableMin} were used for the analysis. Unfortunately, the set of our data is still very
limited and leads to many different results. From these different solutions, we chose to present
here two, which have the lowest rms of the fit. These are given in Table \ref{TableAPSIDAL}. As
one can see, the change of argument of periastron is still rather small and one cannot easily
derive the correct solution. However, the very fast apsidal motion of about 17~years is less
probable due to the typical longer apsidal periods in systems of this type. On the other hand,
the Solution II (see Table \ref{TableAPSIDAL}) presents so slow apsidal motion, that the value
of $\dot\omega = \mathrm{d}{\omega} / \mathrm{d}{t}$ is even so low that the apsidal period $U$
grows to many thousand years. The time spread of epochs of photometry is still too short yet,
hence new precise observations are therefore needed to confirm this hypothesis with higher
conclusiveness.

\begin{table*}[t]
 \scriptsize
 \caption{The parameters of the apsidal motion.}
 \label{TableAPSIDAL} \centering
\begin{tabular}{ c c c }
\hline \hline
 Parameter   &  Value - Solution I           &  Value - Solution II  \\ \hline
 $HJD_0$     & 2454163.2646 $\pm$ 0.0021     & 2454163.2642 $\pm$ 0.0020 \\
 $P$ [day]   & 5.27088632 $\pm$ 0.00000631   & 5.27088780 $\pm$ 0.00000636 \\
 $e$         & 0.0072 $\pm$ 0.0012           & 0.0168 $\pm$ 0.0059 \\
 $\omega_0$ [deg] & 217.5 $\pm$ 1.3          & 251.6 $\pm$ 0.9 \\
 $\dot\omega$ [deg/cycle] & 0.306 $\pm$ 0.013& $<$ 0.0001 \\ \hline
 $U$ [yr]    & 17.1 $\pm$ 0.8                & $>$ 10000 \\
 $P_a$ [day] & 5.27533454 $\pm$ 0.00000632   & 5.27088779 $\pm$ 0.00000636 \\ \hline
\end{tabular}
\end{table*}

\section{Light curve analysis of the system}

For the light curve analysis, there arises several problems with the available data: the photometry
from the "Pi of the sky" survey is unfiltered, and the ASAS data have only poor coverage.
Relatively best data coverage is provided by the SWASP data, but these are not obtained in any
standard photometric filter. The filter used here is a special broadband filter covering a passband
from 400 to 700 nm. Therefore, for the use of these data in our light curve analysis, we used the
filter with the most similar transmission curve, which is the $V_T$ filter of Tycho experiment
onboard of the Hipparcos satellite. The {\sc Phoebe} programme (see e.g. \citealt{Prsa2005}), based
on the Wilson-Devinney algorithm \citep{Wilson1971}, was used for the analysis.

Due to missing information about the stars, and having only the light curve in one filter, some of
the parameters have to be fixed. At first, the "Detached binary" mode (in Wilson \& Devinney mode
2) was assumed for computing. The value of the mass ratio $q$ was set to 1. The limb-darkening
coefficients were interpolated from van~Hamme's tables (see \citealt{vanHamme1993}), the linear
cosine law was used. The values of the gravity brightening and bolometric albedo coefficients were
set at their suggested values for convective atmospheres (see \citealt{Lucy1968}), i.e. $G_1 = G_2
= 0.32$, $A_1 = A_2 = 0.5$. Therefore, the quantities which could be directly calculated from the
light curve are the following: the relative luminosities $L_i$, the temperature of the secondary
$T_2$, the inclination $i$, and the Kopal's modified potentials $\Omega_1$ and $\Omega_2$. The
synchronicity parameters $F_1$ and $F_2$ were also fixed at values of 1.

\begin{figure}
 \includegraphics[width=14cm]{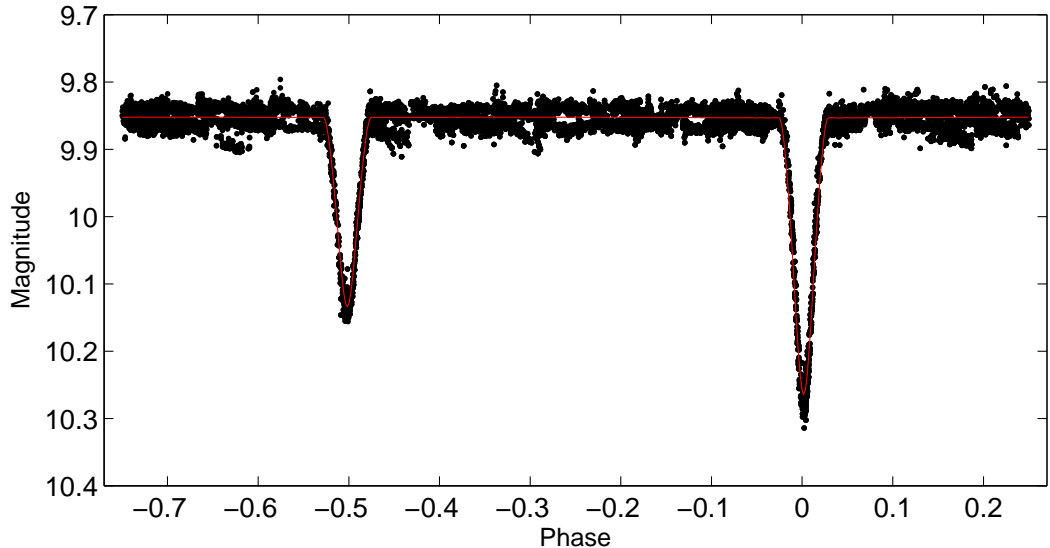}
 \caption{The light curve fit of the SWASP data from {\sc PHOEBE}.}
 \label{FigsLC}
\end{figure}

Because the SWASP data have only limited angular resolution, the third component of the visual
double also influences the photometry and the third light has to be considered. For the light
curve analysis we used the Solution II of the apsidal motion, hence there is no need of apsidal
advance of the $\omega$ angle. The ephemeris were also taken from this solution. The temperature
of the primary component was fixed at a value of 6890~K \citep{2003AJ....125..359W}. Here we
have to emphasize once again that the temperature was only adopted on the basis of a statistical
value for the considered spectral type and not from a good spectroscopic determination. The best
fit we were able to reach is presented in Fig. \ref{FigsLC}. The fit is slightly worse near the
primary minimum, which could suggest the primary minimum is more narrow than the secondary
minimum. The parameters of the fit are given in Table \ref{TableLC}. Regarding the values of M
and R in absolute units, see below.

\begin{table*}[t]
 \scriptsize
 \caption{The light-curve parameters of SY Phe, as derived from our analysis.}
 \label{TableLC} \centering
\begin{tabular}{ c c | c c}
\hline \hline
 Parameter  &  Value             &  Parameter  &  Value       \\ \hline
 $i$ [deg]  & 87.0 $\pm$ 1.3     & $F_1 = F_2$ & 1 (fixed)    \\
 $\Omega_1$ & 12.504 $\pm$ 0.059 & $A_1 = A_2$ & 0.5 (fixed)  \\
 $\Omega_2$ & 12.754 $\pm$ 0.045 & $G_1 = G_2$ & 0.32 (fixed) \\    \cline{3-4}
 $T_1$ [K]  & 6890 (fixed) & \multicolumn{2}{c}{Derived physical quantities:} \\
 $T_2$ [K]  & 6351 $\pm$ 260     & $M_1$ [M$_\odot$] & 1.36 $\pm$ 0.14 \\
 $L_1$ [\%] & 52.1 $\pm$ 1.3     & $M_2$ [M$_\odot$] & 1.36 $\pm$ 0.14 \\
 $L_2$ [\%] & 32.9 $\pm$ 1.0     & $R_1$ [R$_\odot$] & 1.55 $\pm$ 0.09 \\
 $L_3$ [\%] & 15.0 $\pm$ 2.3     & $R_2$ [R$_\odot$] & 1.52 $\pm$ 0.08 \\ \hline
\end{tabular}
\end{table*}

There can be two different explanations of the poor fit to the data. First, the orbital
eccentricity of the binary should be higher, which cause the duration of the primary eclipse
shorter, which is visible in the fit in Fig. \ref{FigsLC}. Other explanation is the fact that there
are some short-periodic variations outside of the eclipses. Therefore, these variations are also
presented in the eclipses and these can shift the data points a bit. The magnitude of the
outside-eclipse variations is up to 0.04~mag. If these variations are physical or instrumental is
still an open question. Moreover, we found that there is also a season-to-season variation of the
light curve. The level of outside of eclipse brightness was changed of about 0.03~mag during the
period of SWASP observations (more than 500~days).

\begin{figure}
 \includegraphics[width=14cm]{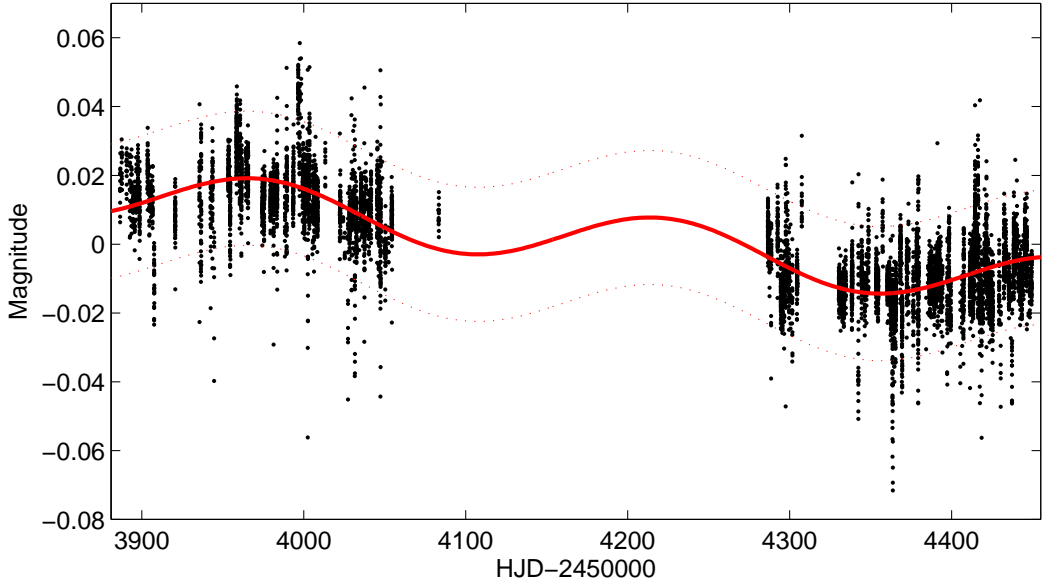}
 \caption{The residuals of the light curve fit with respect to time. The fitted red line represents the fit
 with period of about 249 days and secular decrease of magnitude. Confidence level of 95\% is also shown.}
 \label{FigsResid}
\end{figure}

We were trying to find some periodicity of these data after removing the light curve fit. This
result is shown in Fig. \ref{FigsResid}, where a steady increase of brightness is superimposed with
the sinusoidal variation with the period of about 248.6~days. Having such a long period of
variations, we can only hardly identify these modulations as those making the system classified as
a system "with rapid variations" as stated in Simbad. All of these findings make any analysis
difficult and new more precise standard photometry in different photometric filters would be very
helpful.

\section{Discussion and conclusions}

The first light curve and period analyses of the system SY Phe were performed. Dealing with no
spectroscopy of the target, one has to consider some assumptions and many of the physical
parameters are still affected by relatively large errors.

However, we can roughly estimate the internal structure constants for both apsidal motion
solutions. Assuming the two eclipsing components have masses of about 1.36~M$_\odot$ (spectral type
F4), then the semimajor axis of the orbit is about 17.8~R$_\odot$, which yields the values of radii
of both components (see Table \ref{TableLC}). Using these values, one can calculate the internal
structure constants and compare these values with the theoretical ones, e.g. by
\cite{2004A&A...424..919C}. The Solution I gives $\log$ k$_2$ value of about 0.686, while Solution
II gives $\log$ k$_2$ = -2.198. The latter value agrees well with the theoretical values by
\cite{2004A&A...424..919C}, assuming the F4 star is on main sequence and is about 5 $\cdot 10^8$~yr
old.

The light curve solution also provides the first rough estimation about the third light from the
distant component. Our result of the third light value (15\% of the total light) is in excellent
agreement with the $\Delta$M value provided by the WDS catalogue. The nature of the long-term
photometric variations with period about 249 days still remains an open question. Therefore, new
more detailed analysis is still needed, especially based on new spectroscopic data together with
the photometric observations obtained in various photometric filters.

\section{Acknowledgments}
We thank the "ASAS", "SWASP" and "Pi of the sky" teams for making all of the observations easily
public available. This investigation was supported by the Czech Science Foundation grant no.
P209/10/0715, by the Research Program MSM 0021620860 of the Ministry of Education of Czech
Republic, and by the grant UNCE 12 of the Charles University in Prague. This research has made use
of the SIMBAD database, operated at CDS, Strasbourg, France, and of NASA's Astrophysics Data System
Bibliographic Services.


\begin{thebibliography}{}
 \bibitem[Burd et al.(2005)]{2005NewA...10..409B} Burd, A., et al.\ 2005, NewA, 10, 409
 \bibitem[Claret (2004)]{2004A&A...424..919C} Claret, A.\ 2004, A\&A, 424, 919
 \bibitem[Cox(2000)]{2000asqu.book.....C} Cox, A.~N.\ 2000, Allen's Astrophysical Quantities
 \bibitem[Gim{\'e}nez \& Garc{\'{\i}}a-Pelayo (1983)]{1983Ap&SS..92..203G} Gim{\'e}nez, A., Garc{\'{\i}}a-Pelayo, J.~M.\ 1983, Ap\&SS, 92, 203
 \bibitem[Gim{\'e}nez \& Bastero (1995)]{1995Ap&SS.226...99G} Gim{\'e}nez, A., Bastero, M.\ 1995, Ap\&SS, 226, 99
 \bibitem[H\o g et al.(1997)]{1997A&A...323L..57H} H\o g, E., B{\"a}ssgen, G., Bastian, U., et al.\ 1997, A\&A, 323, L57
 \bibitem[Hoffmeister(1949)]{Hoff1949} Hoffmeister, C.\ 1949, AAAN, 12, Nr 1, 23
 \bibitem[Hoffmeister(1958)]{1958VeSon...3..333H} Hoffmeister, C.\ 1958, Veroeffentlichungen der Sternwarte Sonneberg, 3, 333
 \bibitem[Houk(1978)]{1978mcts.book.....H} Houk, N.\ 1978, Ann Arbor : Dept.~of Astronomy, University of Michigan: distributed by University Microfilms International
 \bibitem[Lucy(1968)]{Lucy1968} Lucy, L.~B.\ 1968, \apj, 151, 1123
 \bibitem[Mason et~al.(2001)]{WDS} Mason, B.~D., Wycoff, G.~L., Hartkopf, W.~I., Douglass, G.~G., Worley, C.~E. 2001, AJ, 122, 3466
 \bibitem[Perryman et al.(1997)]{HIP} Perryman, M.~A.~C., Lindegren, L., Kovalevsky, J., et al.\ 1997, A\&A, 323, L49
 \bibitem[Pojmanski(2002)]{2002AcA....52..397P} Pojmanski, G.\ 2002, AcA, 52, 397
 \bibitem[Pollacco et al.(2006)]{2006PASP..118.1407P} Pollacco, D.~L., et al.\ 2006, PASP, 118, 1407
 \bibitem[Pr{\v s}a \& Zwitter(2005)]{Prsa2005} Pr{\v s}a, A., Zwitter, T.\ 2005, \apj, 628, 42
 \bibitem[Samus et al.(2012)]{GCVS} Samus N.N., Durlevich O.V., Kazarovets E V., et al. General Catalog of Variable Stars (GCVS database, Version 2012Feb)
 \bibitem[Spencer \& Jackson(1936)]{Spencer} Spencer Jones H. Jackson J., 1936, Proper Motions of Stars in the Zone Catalogue -40 to -52 degrees of 20843 Stars for 1900, His Majesty's Stationery Office, London, 1936
 \bibitem[van Hamme(1993)]{vanHamme1993} van Hamme, W.\ 1993, \aj, 106, 2096
 \bibitem[van Leeuwen(2007)]{2007A&A...474..653V} van Leeuwen, F.\ 2007, A\&A, 474, 653
 \bibitem[Wilson \& Devinney(1971)]{Wilson1971} Wilson, R.~E., Devinney, E.~J.\ 1971, \apj, 166, 605
 \bibitem[Wright et al.(2003)]{2003AJ....125..359W} Wright, C.~O., Egan, M.~P., Kraemer, K.~E., \& Price, S.~D.\ 2003, \aj, 125, 359
\end{thebibliography}
\end{document}